\def\year{2020}\relax
\documentclass[letterpaper]{article} % DO NOT CHANGE THIS
\usepackage{aaai20}  % DO NOT CHANGE THIS
\usepackage{times}  % DO NOT CHANGE THIS
\usepackage{helvet} % DO NOT CHANGE THIS
\usepackage{courier}  % DO NOT CHANGE THIS
\usepackage[hyphens]{url}  % DO NOT CHANGE THIS
\usepackage{graphicx} % DO NOT CHANGE THIS
\usepackage{graphicx}  % DO NOT CHANGE THIS
\usepackage{multirow}
\usepackage{tikz}
\usepackage{amsmath}
\usepackage{empheq}

\newcommand{\eat}[1]{}

\title{EGGS: A Flexible Approach to Relational Modeling of Social Network Spam}
\author{
Jonathan Brophy
\and
Daniel Lowd
\\Department of Computer Science
\\University of Oregon
\\\{jbrophy,lowd\}@uoregon.edu
}

\begin{document}

\maketitle

\begin{abstract}
Social networking websites face a constant barrage of spam,
unwanted messages that distract, annoy, and even defraud honest users.
These messages tend to be very short, making them difficult to identify in
isolation. Furthermore, spammers disguise their messages to look
legitimate, tricking users into clicking on links and tricking spam filters
into tolerating their malicious behavior. Thus, some spam filters examine
relational structure in the domain, such as connections among users and
messages, to better identify deceptive content. However, even when it is
used, relational structure is often exploited in an incomplete or ad hoc
manner. 

In this paper, we present Extended Group-based Graphical models for Spam
(EGGS), a general-purpose method for classifying spam in online social
networks. Rather than labeling each message independently, we group related
messages together when they have the same author, the same content, or
other domain-specific connections. To reason about related messages, we
combine two popular methods: stacked graphical learning (SGL) and
probabilistic graphical models (PGM). Both methods capture the idea that
messages are more likely to be spammy when related messages are also
spammy, but they do so in different ways -- SGL uses sequential classifier
predictions and PGMs use probabilistic inference.
We apply our method to four different social network domains. EGGS is more accurate than an independent model in most experimental settings, especially when the correct label is uncertain. For the PGM implementation, we compare Markov logic networks to probabilistic soft logic and find that both work well with neither one dominating, and the combination of SGL and PGMs usually performs better than either on its own.
\end{abstract}

\begin{figure}[ht]
\includegraphics[width=0.475\textwidth]{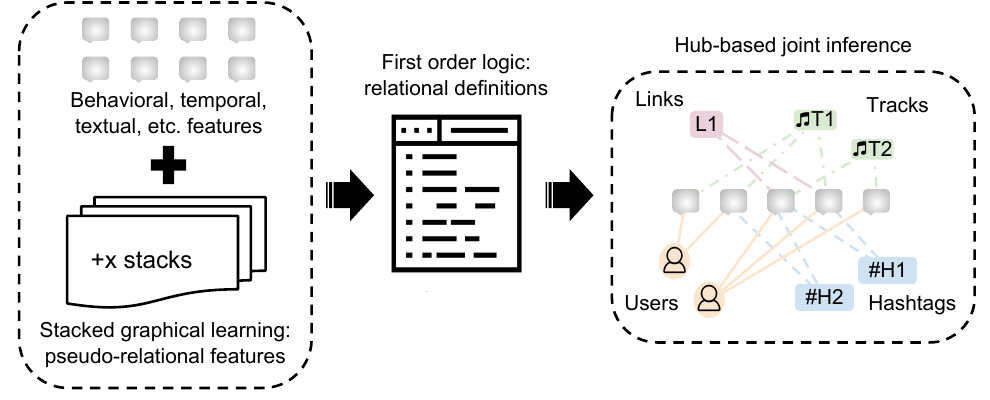}
\caption{EGGS workflow, with example relational rules (right).}
\label{fig:label_prop}
\end{figure}

\section{Introduction}

Social spam~\cite{chu2012detecting} is any unsolicited or unwanted action by a user in a social network. Many methods have been developed to detect spam based on the content of the messages themselves~\cite{blanzieri2008survey,costa2013defining,ferragina2015analyzing,liu2016detecting,ma2013predicting,7301005,xu2016efficient,zhang2012detecting}, the graph structure
among users and messages~\cite{akoglu2013opinion,boykin2005leveraging,fakhraei2015collective,ghosh2012understanding,o2012network,rayana2015collective}, the timing of user
actions~\cite{fei2013exploiting,stringhini2010detecting,tan2013unik,viswanath2014towards,7301005,zhu2012discovering}, and more. 
This works well when there are clear patterns that distinguish spam from non-spam, but can fail when spammers obfuscate their behavior.
A complementary approach is to exploit relationships among different users and messages, so that known spammers and spam can be used to identify other spammers and spam~\cite{akoglu2013opinion,fakhraei2015collective,li2014spotting,rayana2015collective,laorden2011collective,chen2009co,wu2015social}. This works well when there are strong predictive relationships linking entities, such as textual similarity among messages or friendship in a social network, but can fail when spammers start new campaigns that are not connected to previously known ones.

In this paper, we integrate these ideas into a flexible method for classifying social spam: Extended Group-based Graphical models for Spam (EGGS). EGGS begins with predictions from an independent classifier, which can use any number of domain-appropriate features. To incorporate relationships among different messages, EGGS defines groups of related messages that have the same author, the same text, or other domain-specific similarities such as the same hashtags. A message can be in multiple groups representing different types of relationships. Related messages are more likely to have the same label, although this probability may depend on the type of relationship. EGGS models these relationships with a probabilistic graphical model, using one of four different approaches: stacked graphical learning (SGL)~\cite{kou2007stacked}, Markov logic networks~(MLNs)~\cite{richardson2006markov}, probabilistic soft logic~(PSL)~\cite{bach2015hinge}, or a new combination of SGL with either MLNs or PSL~(Figure~\ref{fig:label_prop}). We show that this integrated approach can accurately detect spam on large real-world datasets for multiple domains, requiring very few modifications.

In spite of the breadth of prior work in this area, most other methods fall short in one of three ways:
\begin{itemize}

\item They're \emph{specialized for another domain}, such as detecting fake reviews or auction fraud. Social spam has its own structures which are distinct from other adversarial domains. Nonetheless, we can still generalize among different social spam domains. In Section~\ref{sec:methodology}, we introduce a general framework for social spam and apply it to SoundCloud, Twitter, YouTube, and Wikipedia.

\item They \emph{ignore content and other features}, and only use the relational structure plus a small number of known labels. In Section~\ref{sec:evaluation}, we show that new social spam is often unconnected to previously seen spam, and that social spam forms many distinct connected components. This means that labels on the training data or a small number of labels in the test data will not cover most spam.
\item They \emph{ignore relational structure} and make predictions for each message independently. Our empirical results show that exploiting relational structure can lead to substantially improved performance over independent predictions.
\end{itemize}

Our primary contributions are EGGS, a flexible method for social network spam that overcomes all three of the above limitations, and the application and evaluation of EGGS on three different domains. As secondary contributions, we show that stacked graphical learning can be combined with other probabilistic models for improved performance, and that relational modeling can scale to large domains using simple methods.

\section{Related Work}\label{sec:related_work}

We give a brief overview of the many techniques used to detect spam from both independent and relational perspectives.

\subsection*{Spam Detection}

Much of the work done on social spam filtering involves an analysis of some narrow set of specific independent features. Bag-of-words models have been popular ever since the rise of email spam, and continue to see use in Youtube and Twitter models~\cite{song2014discriminative,mateen2017hybrid,xu2016efficient}. Word embeddings~\cite{savigny2017emotion} and LDA topic modeling~\cite{song2014discriminative} have also been effective. Numerous works focus on URLs~\cite{benevenuto2010detecting,mateen2017hybrid,ghosh2012understanding,7301005,gao2012towards,stringhini2010detecting,wang2010don,chu2012detecting}, while others look at hashtags~\cite{mateen2017hybrid,ferragina2015analyzing,ma2013predicting,sedhai2015hspam14,wang2010don,chu2012detecting} and mentions~\cite{mateen2017hybrid,wang2010don,chu2012detecting}. Each of these have unique advantages, but they can all be seen as content-related features, derived directly from the messages themselves.

User-based features attempt to characterize the behavior of a user in the network with the hope that this will distinguish normal users from malicious ones. A popular approach is to capture the `burstiness' of user activity~\cite{fei2013exploiting,gao2012towards,kc2016temporal,li2017bimodal,song2014discriminative,viswanath2014towards}. Other approaches look at the number of follows~\cite{mateen2017hybrid,benevenuto2010detecting}, the types and sequences of user actions~\cite{fakhraei2015collective,viswanath2014towards}, the ratio of different types of messages users send~\cite{gao2012towards}, and account profiles~\cite{mateen2017hybrid,chu2012detecting}. Graph-based features are conceptually a subset of user-based features, and are derived specifically from a directed graph built using user interactions (e.g. one user following another) on which graph features such as pagerank, betweenness, in/out degree, etc.\ are computed~\cite{mateen2017hybrid,fakhraei2015collective,gao2012towards,7301005,wang2010don}. Similarly, works such as \emph{CopyCatch}~\cite{beutel2013copycatch}, \emph{CatchSync}~\cite{jiang2014catchsync}, \emph{Fraudar}~\cite{hooi2016fraudar}, and \emph{Gang}~\cite{wang2010don} use the subgraph density of the graph topology to detect anomalous behavior.

Many of these works use a combination of approaches mentioned above, and Mateen et al.~\cite{mateen2017hybrid} investigate a more comprehensive hybrid approach combining features from all three categories to detect spam on Twitter.

\subsection*{Collective Filtering}

The works by Pandit et al. and Akoglu et al.~\cite{pandit2007netprobe,akoglu2013opinion}, \textit{NetProbe} and \textit{FraudEagle}, use Markov random fields~(MRFs) with belief propagation~(BP) to detect fraudsters on eBay and the Software Market App Store using only the network structure of users and products, with the edges between them representing positive or negative reviews~\cite{jindal2007analyzing}; Akoglu et al.~\cite{rayana2015collective} expand upon this with \textit{SpEagle}, generating unsupervised priors for users and products to propagate in the network structure. Li et al.~\cite{li2014spotting} use iterative classification~(ICA) between review, user, and IP nodes to detect spammers on Dianping, a network similar to Yelp. These are all unsupervised methods, which are especially useful for detecting opinion spam~\cite{jindal2007analyzing} because true labels are often difficult to obtain for this problem domain; this has the added benefit of not having to train an additional supervised classifier on separate data to generate priors to propagate. For social spam, labeled training data is more prevalent, allowing us to build more powerful priors using supervised classifiers, which are then propagated by our relational methods.

Works on social spam, such as Duan et al.~\cite{duan2012graph}, experiment using ICA, BP, and relaxation labeling on the message-message graph connected by similar URLs and hashtags to classify topics on a small Twitter dataset. Li et al.~\cite{li2014detecting} use typed MRFs, connecting users with URLs and tweet bursts to spot spam campaign promoters on Twitter. Fakhraei et al.~\cite{fakhraei2015collective} leverage user reports using PSL to find spammers in the on-line dating network IfWe; however, the relational rules were only relevant to a small proportion of users, reducing the effectiveness of joint reasoning. Castillo et al.~\cite{castillo2007know} use stacked graphical learning~(SGL) to introduce one new relational feature in addition to their original features to better detect spam hosts on the \textit{Webspam-UK2006} dataset.

Matrix factorization methods are a complementary approach. Zhu et al.~\cite{zhu2012discovering} encode user-user relations from different user interaction types using collective matrix factorization~\cite{singh2008relational} to detect spammers on RenRen. Shen et al.~\cite{shen2015detecting} extend this by adding a social interaction coefficient to spot spammers on Twitter. Chen et al.~\cite{chen2009co} tackle the spam and spammer problems simultaneously using relations between users and bookmarks on the website: delicious.com. Wu et al.~\cite{wu2015social} take the same approach but use different relations (user-user, user-message, and message-message connected by URLs and hashtags) to work on the platform Sina Weibo.

\section{Methodology}\label{sec:methodology}

We now introduce Extended Group-based Graphical models for Spam (EGGS), our framework for detecting social network spam. The basic approach is to predict the label of each message using standard classification methods, and then refine those predictions using relational reasoning methods on groups of related messages. EGGS is not a single, monolithic method, but a general approach that can incorporate any domain-specific features and relations, any type of classifier, and any type of probabilistic graphical model. In the following, we describe a set of features and methods that work well for detecting social network spam in several domains.

\subsection{Independent Modeling}\label{sec:independent_model}

We begin with a standard classifier, which we refer to as the ``independent model,'' since it makes predictions for each message separately.
We highlight engineered features shown to work well for spam classification in social networks based on previous research. In addition to the features listed here, it is easy to extend our method to add others for different domains.

\paragraph{Content-based Features}
We use the text of each message to generate content-based features such as: the number of characters, hashtags, links, and the top 10,000 tri-grams (selected based on term frequency) used as binary features~(Table~\ref{tab:ind_feats}:~Content).

\paragraph{User-based Features}
We aggregate user actions (although any relevant entity in the domain may be used: users, URLs, tracks, videos, hashtags) to create user-based features~(Table~\ref{tab:ind_feats}:~User). Features are computed in sequential order of messages based on their timestamp. For example, when computing the feature \textit{UserUploads} for message \#100 posted by user x, we record the number of tracks uploaded by user x up until message \#100. This creates a more realistic scenario as features are computed only based on previous messages.

\paragraph{Graph-based Features}
As done in prior work (e.g.,~\cite{fakhraei2015collective}), we create graph features using a list of follower actions. We can represent this list of affiliations as a graph, where we construct a node for every user in the list, and add a directed edge from user $x$ to user $y$ whenever user $x$ starts to follow user $y$. Then we compute the following features on the resulting graph: \emph{Pagerank}~\cite{page1999pagerank}, \emph{Triangle count}~\cite{schank2005finding}, \emph{K-core}~\cite{alvarez2006large},
\emph{In/Out-degree}~\cite{newman2001random}~(Table~\ref{tab:ind_feats}:~Graph). Each feature represents a different aspect of connectivity for a user to the community, and capitalizes on the assumption that spammers tend to be less connected than non-spammers~\cite{fakhraei2015collective}, or more connected with other suspicious users.

\begin{table}[!ht]
\footnotesize
\caption{Independent Model Features}
\label{tab:ind_feats}
\centering
\begin{tabular}{ll}
\hline
\noalign{\vskip 1mm}
\textbf{Content} & \\
\noalign{\vskip 1mm}
\textit{NumChars} & \# chars in msg~(Sanchez et al.~\citeyear{sanchez2011twitter}) \\
\textit{NumHashtags} & \# hashtags in msg~\cite{mateen2017hybrid} \\
\textit{NumLinks} & \# links in msg~\cite{gao2012towards}. \\
\textit{NumMentions} & \# mentions in msg~(Chu et al.~\citeyear{chu2012detecting}). \\
\textit{IsRetweet} & 1 if msg is a retweet~\cite{jiang2011target}. \\
\textit{Polarity} & Msg. polarity~(Sanchez et al.~\citeyear{sanchez2011twitter}). \\
\textit{Subjectivity} & Msg. subjectivity~(Sanchez et al.~\citeyear{sanchez2011twitter}). \\
\textit{N-grams} & Top 10,000 tri-grams~(Song et al.~\citeyear{song2014discriminative}). \\

\hline
\noalign{\vskip 1mm}
\textbf{User} & \\
\noalign{\vskip 1mm}
\textit{UMsgs} & \# msgs posted per user~\cite{mateen2017hybrid}. \\
\textit{UHRatio} & Frac. user msgs w/ a \#~\cite{mateen2017hybrid}. \\
\textit{UMRatio} & Frac. user msgs w/ a @~(Chu et al.~\citeyear{chu2012detecting}). \\
\textit{ULRatio} & Frac. user msgs w/ a URL~\cite{gao2012towards}. \\
\textit{UBlacklist} & 1 if user posts 3+ spam~(Chu et al.~\citeyear{chu2012detecting}). \\
\textit{UWhitelist} & 1 if user posts 10+ ham~(Chu et al.~\citeyear{chu2012detecting}). \\
\textit{UMsgMax} & Max msg len by user~(Sanchez et al.~\citeyear{sanchez2011twitter}). \\
\textit{UMsgMin} & Min msg len by user~(Sanchez et al.~\citeyear{sanchez2011twitter}). \\
\textit{UMsgMean} & Avg msg len by user~(Sanchez et al.~\citeyear{sanchez2011twitter}). \\
\textit{TMsgs} & \# msgs per track. \\

\hline
\noalign{\vskip 1mm}
\textbf{Graph} &  \\ 
\noalign{\vskip 1mm}
\textit{Pagerank} & Pagerank of each node in the \\
 & follower graph G~\cite{page1999pagerank}. \\
\textit{TriCnt} & \# triangles per node~(Schank et al.~\citeyear{schank2005finding}). \\
\textit{KCore} & Iteration node is \\
& pruned~(Alvarez et al~\citeyear{alvarez2006large}). \\
\textit{InDegree} & \# edges entering node~(Newman et al.~\citeyear{newman2001random}). \\
\textit{OutDegree} & \# edges leaving node~(Newman et al.~\citeyear{newman2001random}). \\

\end{tabular}
\end{table}

\begin{table}[!ht]
\footnotesize
\caption{Stacking Features}
\label{tab:stacking_feats}
\centering
\begin{tabular}{ll}
\hline
\noalign{\vskip 1mm}
\textit{MMSRatio} & Fraction of spammy matching messages. \\
\textit{USRatio} & Fraction of spammy messages per user. \\
\textit{TSRatio} & Fraction of spammy messages per track. \\
\textit{HSRatio} & Fraction of spammy messages per hashtag. \\
\textit{MSRatio} & Fraction of spammy messages per mention. \\
\textit{LSRatio} & Fraction of spammy messages per link. \\
\textit{HSRatio} & Fraction of spammy messages per user hashtag. \\
\end{tabular}
\end{table}

\subsection{Relational Modeling}\label{sec:pseudo-relational}

EGGS uses several types of relational modeling, separately or in combination, to improve on the independent model.

\subsubsection{Stacked Graphical Learning}

SGL~\cite{kou2007stacked} is a simple approach for performing collective classification, in which the label of each entity depends on the labels of its neighbors. Since the true labels of the neighbors are often unknown, SGL first uses an independent classifier to predict the label of every entity. The predicted labels can then be used to derive features for a second classifier, which makes a refined prediction for each entity. This process can be repeated multiple times, ``stacking'' classifiers on top of classifiers to any depth. 

\begin{figure}[t]
\begin{center}
\includegraphics[width=0.25\textwidth]{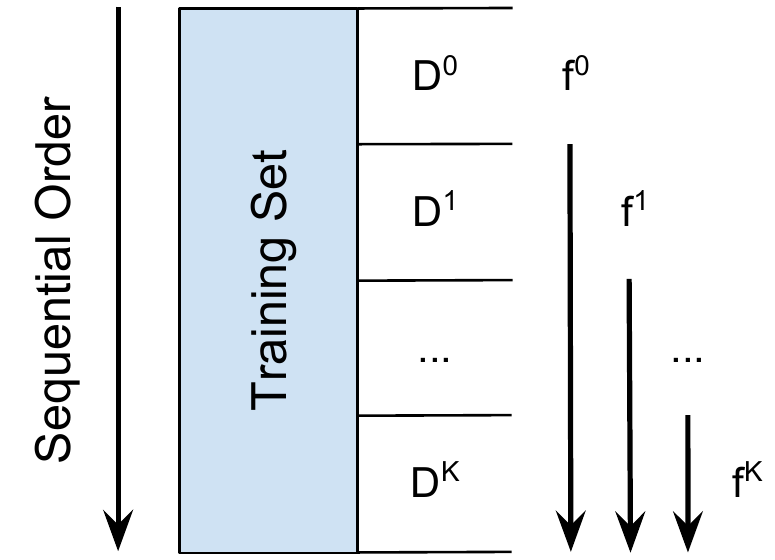}
\caption{Stacked graphical learning with K stacks using the holdout method.}
\label{fig:stacking}
\end{center}
\end{figure}

We apply this idea to social network spam by defining ``pseudo-relational'' features, each of which summarizes the predicted labels of related messages (Table~\ref{tab:stacking_feats}). For example, ``USRatio'' is the fraction of messages written by the same user which are predicted to be spam. Since most spammers send multiple spam messages, a higher value of USRatio indicates that the message is more likely to be spam. We create pseudo-relational features based on each relation, contrary to the previous application of SGL on webspam where they create only one additional feature: the average predicted label of its neighbors. The learning procedure is similar to previous approaches~\cite{castillo2007know,kou2007stacked}, except we use a holdout method instead of cross-validation when building the sub-model for each stack~(Figure~\ref{fig:stacking}).

Using K stacks, we evenly split the training data $D=\bigcup_k^KD^k$, and train a base model $f^0$ on $D^0=(X_g^0, y^0)$, where $X_g^0$ are the original features for $D^0$. Each subsequent submodel $f^k$ trains on $D^k$ with additional pseudo-relational features:
\[
f^k = 
\begin{cases} 
      train(X_g^k, y^k)  & k = 0 \\
      train(X_g^k + X_{p_{k-1}}^k, y^k) & k > 1 \\
   \end{cases}
\]
$X_{p_k}^j$ are pseudo-relational features for $D^j$ using predictions from $f^k$, which we can define as a function of $X_g^j$ and $\hat{y}_k^j$ (predictions for $D^j$ from $f^k$):
\begin{equation*}
X_{p_k}^j = S(X_g^j, \hat{y}_k^j)
\end{equation*}
where
\[
\hat{y}_k^j = 
\begin{cases}
      f^k(X_g^j)  & j > 0, k = 0 \\
      f^k(X_g^j + S(X_g^j, \hat{y}_{k-1}^j)) & j > 1, k > 0, k < j \\
   \end{cases}
\]
Then given test data, $X'$, inference is the same as the original SGL~\cite{kou2007stacked}: 
\vspace{0.1in} \\
\begin{tabular}{l}
$\hat{y}'_0 = f^0(X'_g)$ \\
For k = 1...K: \\
~~~~~~~~$X'_{p_k} = S(X'_g, \hat{y}'_{k-1})$ \\
~~~~~~~~$\hat{y}'_k = f^k(X'_g + X'_{p_k})$ \\
return $\hat{y}'_k$
\end{tabular}
\vspace{0.1in}

This holdout method builds submodels in sequential order of the data, while also computing the pseudo-relational features in sequential order. This makes the problem more realistic since we do not ignore the temporal component of the data, contrary to the cross-validation technique. However, this comes with a cost; the cross-validation technique~\cite{castillo2007know,kou2007stacked} trains each submodel on the entire training set. Thus, more stacks generally does not decrease performance. In our case, the higher K becomes, the less data each submodel has to train on, introducing the possibility of underfit models. In practice, we find that 1-2 stacks works best, which is consistent with previous works~\cite{castillo2007know,kou2007stacked,fast2008stacked}.

\subsubsection{Flexible Joint Inference}\label{sec:relational_model}

Statistical relational learning (SRL) has developed various methods for doing collective classification~\cite{getoor2007introduction}, most promisingly using probabilistic graphical models (PGMs).
We experiment with Markov logic networks (MLNs)~\cite{richardson2006markov} and probabilistic soft logic (PSL)~\cite{bach2015hinge}, both of which use weighted formulas in first-order logic to define a template for a PGM. 
Like stacking, the goal is to improve the predicted label for each message by reasoning about the labels of related messages. However, instead of a sequential pipeline of classifiers, MLNs and PSL define a joint distribution over all possible labelings and perform probabilistic inference to reason about that distribution.

We use the formulas to capture our intuition about how information should propagate among related messages, and then describe how we use Markov logic and PSL to turn these formulas into a full probabilistic model. Our relational model contains two main components: priors (negative and positive) and relations. The negative prior assumes that all messages are non-spam, while the positive prior gives the model information to propagate~(Figure~\ref{rel_model}~(a,b)). We focus on connecting messages to one another using any relation, but our model may connect entities of any type together, such as users, hashtags, URLs, etc.

\begin{figure}[tb]
\begin{empheq}[box=\fbox]{align*}
&\neg spam(e) \tag{a} \\
prior(e) \rightarrow& \ spam(e) \tag{b} \\
hasRel(r, e) \wedge spam(e) \rightarrow& \ spamRel(r) \tag{c} \\
hasRel(r, e) \wedge spamRel(r) \rightarrow& \ spam(e) \tag{d}
\end{empheq}
\caption{Generalized relational model represented as first order logical rules. (a) Negative prior, (b) positive prior, (c, d) hub-based relational rules; e is short for entity (typically messages or users), r is short for relation.}
\label{rel_model}
\end{figure}

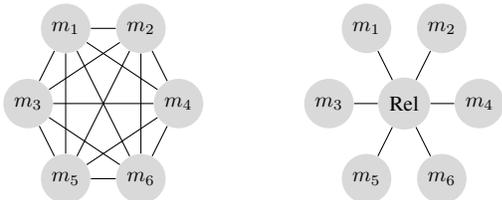
\begin{figure}
\centering
\begin{tikzpicture}

% pairwise drawing
\node [circle, fill=gray!30, scale=0.8] (A) at (1.5,2) {$m_1$};
\node [circle, fill=gray!30, scale=0.8] (B) at (2.5,2) {$m_2$};
\node [circle, fill=gray!30, scale=0.8] (C) at (1,1) {$m_3$};
\node [circle, fill=gray!30, scale=0.8] (D) at (3,1) {$m_4$};
\node [circle, fill=gray!30, scale=0.8] (E) at (1.5,0) {$m_5$};
\node [circle, fill=gray!30, scale=0.8] (F) at (2.5,0) {$m_6$};
\draw (A) -- (B) -- (D) -- (F) -- (E) -- (C) -- (A);
\draw (A) -- (D) -- (E) -- (B) -- (C) -- (F) -- (A);
\draw (A) -- (E); \draw (B) -- (F); \draw (C) -- (D);

% hub-based drawing
\node [circle, fill=gray!30, scale=0.8] (G) at (5.5,2) {$m_1$};
\node [circle, fill=gray!30, scale=0.8] (H) at (6.5,2) {$m_2$};
\node [circle, fill=gray!30, scale=0.8] (I) at (5,1) {$m_3$};
\node [circle, fill=gray!30, scale=0.8] (J) at (7,1) {$m_4$};
\node [circle, fill=gray!30, scale=0.8] (K) at (5.5,0) {$m_5$};
\node [circle, fill=gray!30, scale=0.8] (L) at (6.5,0) {$m_6$};
\node [circle, fill=gray!30, scale=0.8] (M) at (6,1) {Rel};
\draw (G) -- (M); \draw (H) -- (M); \draw (I) -- (M);
\draw (J) -- (M); \draw (K) -- (M); \draw (L) -- (M);
\end{tikzpicture}
\caption{Pairwise connections between related messages~(left). Hub approach connecting related messages~(right).}
\label{fig:connections}
\end{figure}

The second part defines the relations to exploit from the network structure, grouping related messages together~(Figure~\ref{rel_model}~(c,d)). Previous work using Markov networks for joint reasoning over a set of related entities typically do so in a direct fashion~\cite{tan2013unik}. For example, we could have written rules (c) and (d) as:
\begin{align}
spamRel(e_1, e_2) \wedge spam(e_1) \rightarrow spam(e_2)
\end{align}
\noindent This type of modeling provides propagation directly from one message to another, but makes it difficult to scale for large groups of related entities. For example, a group of 100 related messages would need $\binom{100}{2}$ edges to take care of every possible interaction. Our concept of a `hub node' essentially creates a hyper-edge from one entity to all related entities~(Figure~\ref{fig:connections}). This significantly reduces the number of edges to be linear in the number of related entities. As information starts to propagate, these `hub nodes' become more or less `spammy', which in turn propagates to entities that are difficult to detect at first, but become clear as their connections to other entities are exploited.

We use the outputs of the supervised classifier as positive priors for our joint prediction model, but it is important to note that a message with no relations to any other messages in the dataset will not be affected by this model.

\paragraph{MLN Implementation}

We first implement our relational model as an MLN, and then convert our MLN into an MRF using the Libra toolkit~\cite{JMLR:v16:lowd15a} since belief propagation in Libra is better optimized than inference in existing MLN implementations. We can convert our MLN formulas into equivalent MRF factor potentials between a message node and a hub node~(Table~\ref{tab:mrf_relations}), where $\epsilon$ is tuned separately for each relation and inference is done using loopy belief propagation, similar to prior work on fraud detection~\cite{pandit2007netprobe,rayana2015collective}.

\begin{table}[ht!]
\caption{Factor Potential Definition per Relation}
\label{tab:mrf_relations}
\footnotesize
\centering
\begin{tabular}{|l|c|c|}
\hline
& \multicolumn{2}{c|}{\textbf{Message State}} \\
\hline
\textbf{Hub State} & spam & non-spam \\
\hline
spam & 1 - $\epsilon$ & $\epsilon$ \\
\hline
non-spam & $\epsilon$ & 1 - $\epsilon$ \\
\hline
\end{tabular}
\end{table}

\paragraph{PSL Implementation}

The second implementation of our relational rules is as a PSL model, which builds a hinge-loss Markov random field (HL-MRF)\cite{bach2015hinge}: a type of log-linear model that uses hinge loss functions of the variable states as features and can be modeled as a conditional probability distribution as follows~\cite{fakhraei2015collective}:
\begin{align}
P(Y|X) = \frac{1}{Z(\omega)}\exp\Big(-\sum_{i=1}^n \omega_i \phi_i(X,Y)\Big)
\label{cond_prob_formula}
\end{align}
\noindent where $\phi_i$ is the set of $n$ continuous potentials:
\begin{align}\label{dist_to_satisfaction_formula}
\phi_i(X,Y) = [\max\{0, \ell_i(X,Y)\}]^{p_j},
\end{align}
\noindent $\ell$ is a linear function of X and Y, and $p_j \in$~\{1,2\}. We can learn the weights to these rules from data using gradient descent and expectation maximization~\cite{bach2015hinge}.

Both of these models propagate information among the same sets of related messages, but there is an important difference: loopy belief propagation in MLNs can combine uncertain prior probabilities to arrive at more confident posterior probabilities. For example, if $n$ messages all the have the same prior of 0.85, their posterior scores can be `pushed' beyond 0.85, and this effect increases as $n$ increases. On the contrary, spam scores in the PSL model stop increasing once their distance to satisfaction specified in~(\ref{dist_to_satisfaction_formula}) reaches zero.

\section{Evaluation}\label{sec:evaluation}

We evaluate the effectiveness of EGGS at finding spam in four social network domains. We use the area under the precision-recall curve (AUPR) as our main metric. AUPR measures the ability to find most of the positive examples (spam) without too many negative examples (non-spam). With highly imbalanced classes, as in our SoundCloud dataset (1.6\% spam), AUPR is usually a better way to differentiate between classifiers than alternatives such as area under the ROC curve.

\subsection{Spam Detection}\label{sec:spam_detection}

We evaluate our methods on spam data from three social networks: SoundCloud, YouTube, and Twitter. SoundCloud is an online music sharing network where users can upload original tracks that other users can listen to and comment on. This dataset includes all comments posted from 10/10/2012 to 9/30/2013 on approximately 8M tracks~(Table~\ref{tab:data}). YouTube is a video sharing service similar to SoundCloud. The data\footnote{http://mlg.ucd.ie/yt/} was collected from 10/31/2011 to 1/17/2012 with a focus on the most viewed and top-rated videos~\cite{o2012network}. This dataset contains no user subscription attributes (analogous to followers in SoundCloud and Twitter), as the data collectors were often restricted from this information~\cite{o2012network}. We use the Twitter \emph{HSpam14 Dataset}\footnote{http://www.ntu.edu.sg/home/axsun/datasets.html}, curated from 1/5/2013 to 6/31/2013 with a hashtag-oriented focus~\cite{sedhai2015hspam14}. \\

\begin{table}
\caption{Basic Statistics per Domain}
\label{tab:data}
\centering
\resizebox{0.47\textwidth}{!}{
\begin{tabular}{lrrr}
\hline
\noalign{\vskip 1mm}
\textbf{Entity} & \textbf{SoundCloud} & \textbf{YouTube} & \textbf{Twitter} \\
\hline
\noalign{\vskip 1mm}
messages & 42,783,305 & 6,431,471 & 8,845,979 \\
spam & 684,338 & 481,334 & 1,722,144 \\
users & 5,505,634 & 2,860,264 & 4,831,679 \\
spammers & 128,016 & 177,542 & 843,002 \\
follows & 335,000,000 & N/A & 128,000,000 \\
\end{tabular}}
\end{table}

\begin{figure*}[ht]
  \mbox{\includegraphics[width=0.33\textwidth]{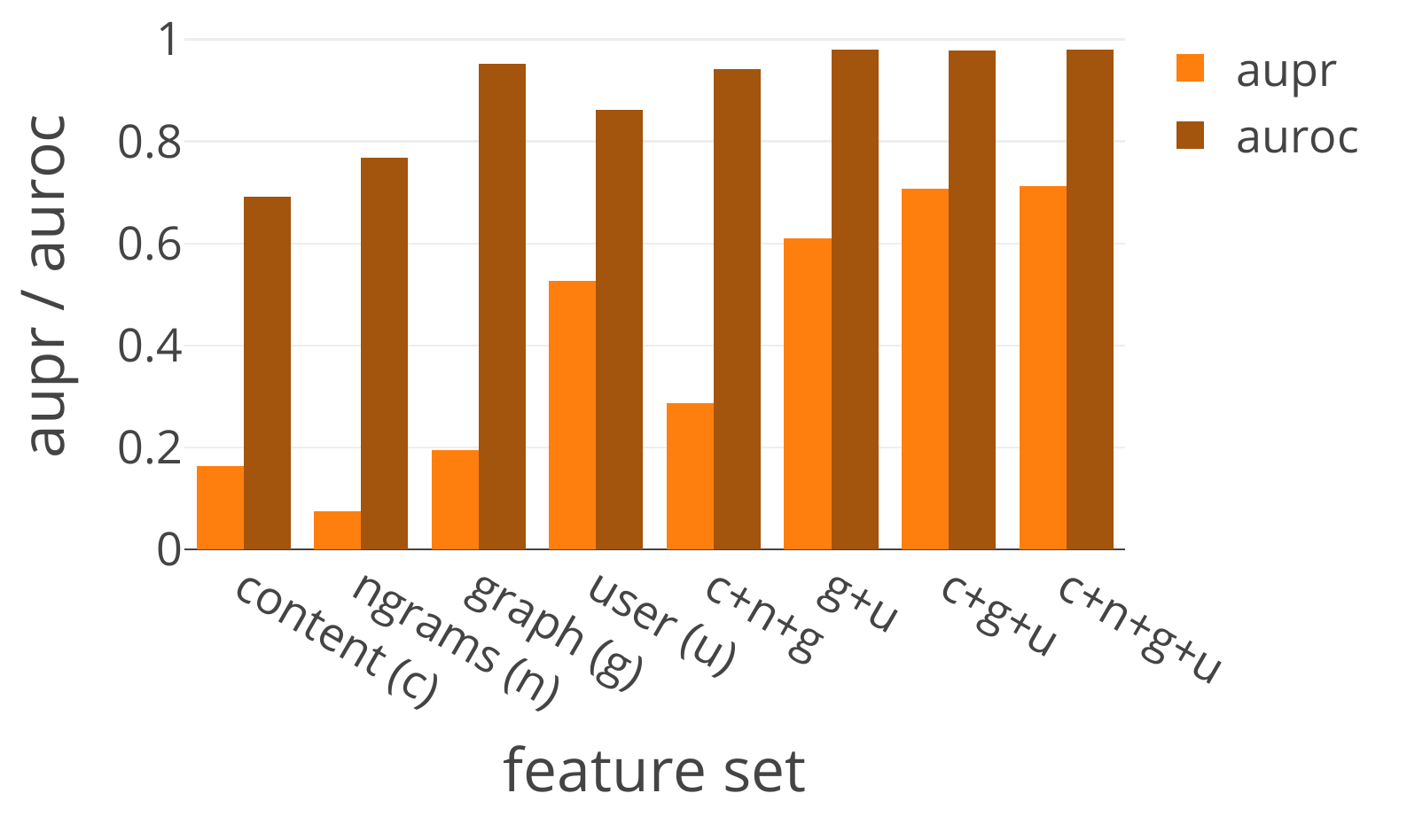}}
  \mbox{\includegraphics[width=0.33\textwidth]{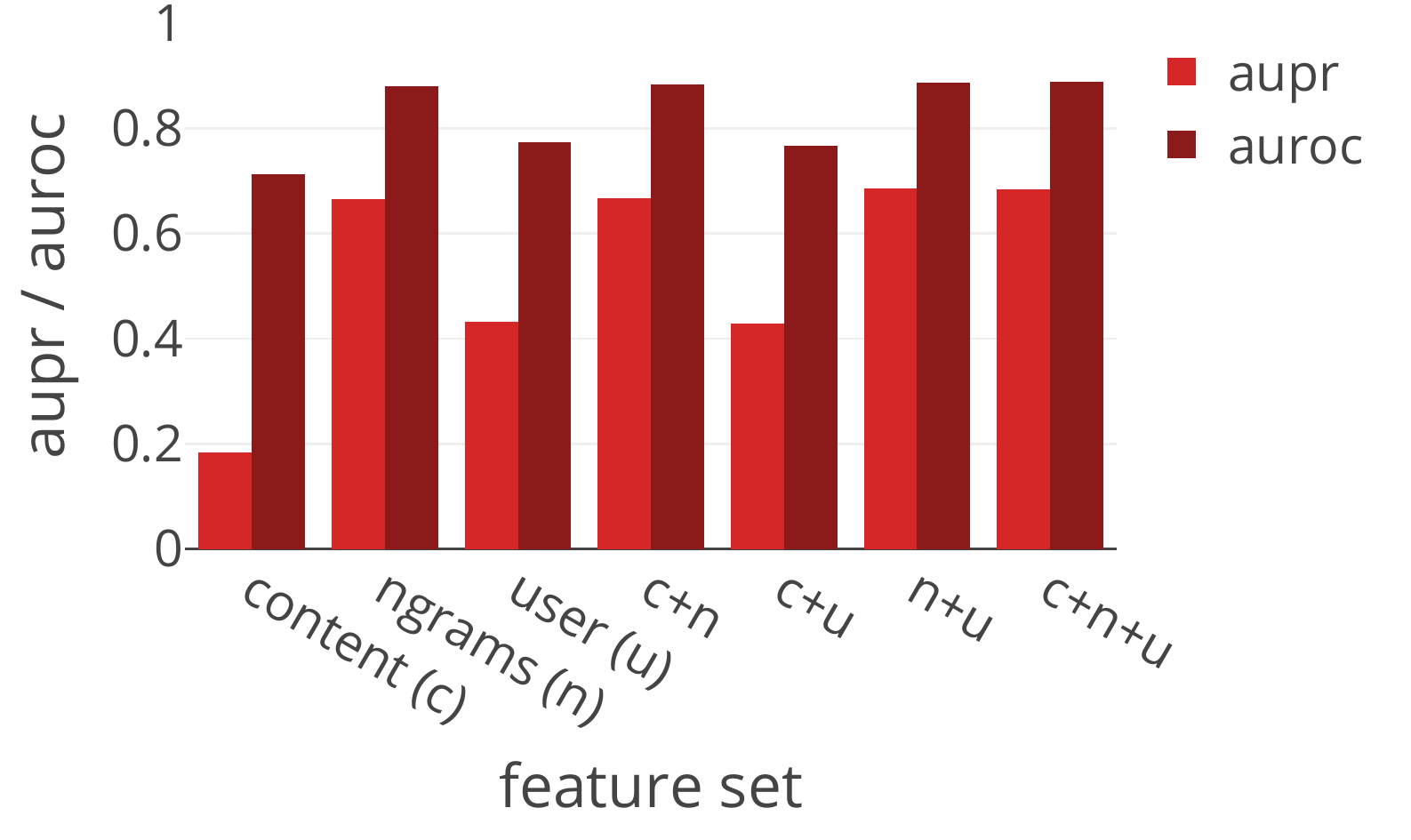}}
  \mbox{\includegraphics[width=0.33\textwidth]{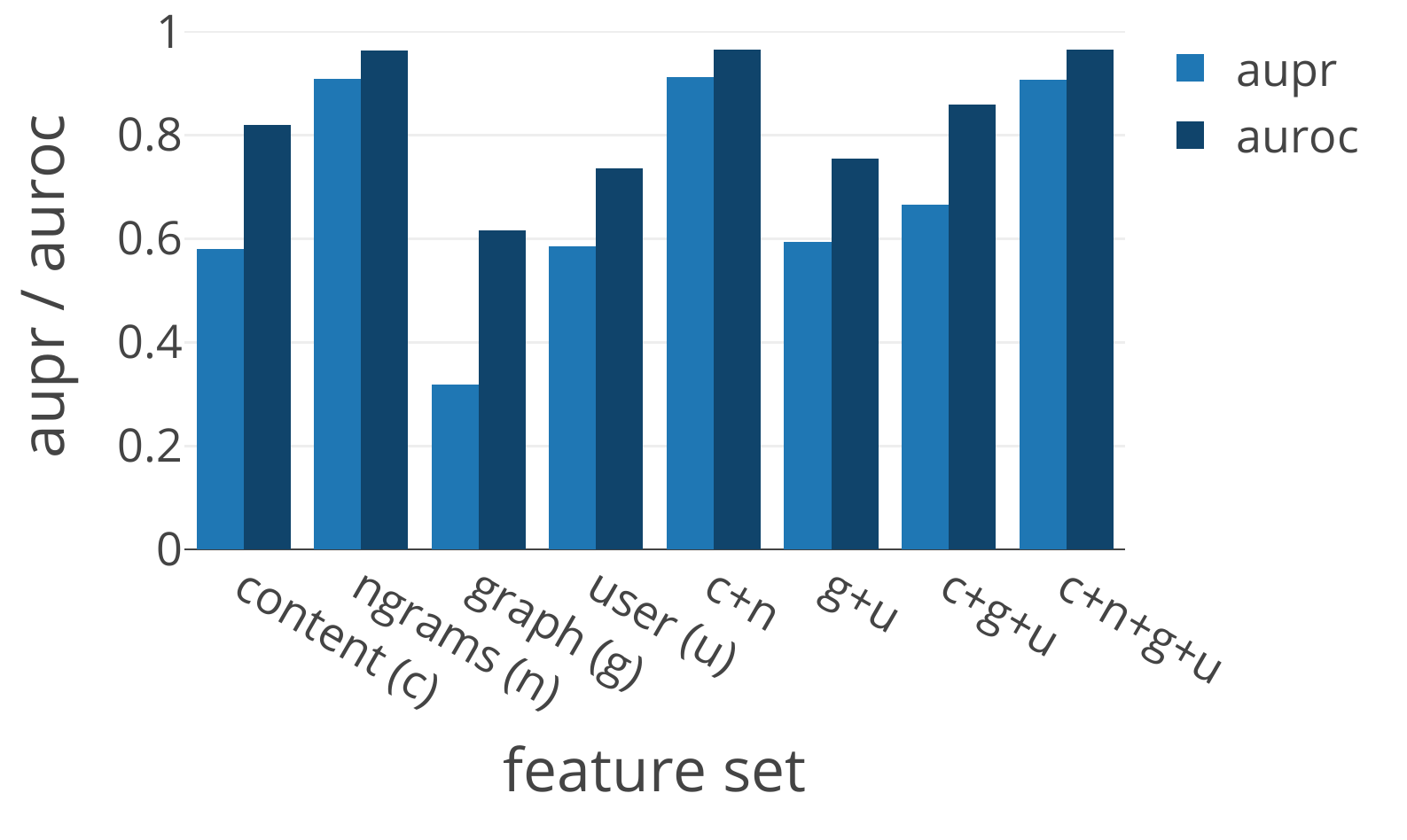}}
  \caption{Feature set performance for SoundCloud (left), YouTube (middle), and Twitter (right).}
  \label{fig:ablation}
\end{figure*}

Social network spam evolves rapidly, so a method that works well one month may work poorly the next. We incorporate this into our experiments by creating ten test sets for each domain, partitioned chronologically. Predictions from each test set are concatenated into one large test set, on which the AUPR is computed. Different test sets may reflect different spam campaigns, user policy changes for the network, and changing usage patterns among legitimate users over time. For each test set, we use data from the preceding time period for training\footnote{For the Twitter dataset, time information was not available, so the tweets were ordered based on tweet ID.}. This ensures that each prediction is only made using data from the past, never the future.

SoundCloud is evenly split into ten non-overlapping subsets of roughly 4M messages each (70\% training, 1.25\% validation, and 28.75\% testing). We exploit the following relations between messages: \emph{users} (messages posted by the same user), \emph{text} (messages with similar text), and \emph{links} (messages with similar links). YouTube is split into ten overlapping subsets of roughly 2M messages each (75\% training, 2.5\% validation, and 28.5\% testing). Since this dataset only contains 6.4M messages, these subsets contain some overlap, but the test sets remain mutually exclusive, and we use the following relations: \emph{users} and \emph{text}. Twitter is evenly split into ten non-overlapping subsets of roughly 880K messages (70\% training, 6\% validation, and 24\% testing); we use the following relations: \emph{users}, \emph{text}, and \emph{user hashtags} (hashtags posted by a given user). We learn the weights for our PSL model on the validation data, and pre-tune $\epsilon$ in the MRF model for each relation.

\begin{table*}[ht]
\caption{Spam Detection Performance (AUPR)}
\label{tab:overall_performance}
\begin{center}
\resizebox{!}{3.5cm}{
\begin{tabular}{|cc||ccc||ccc|}
\hline
& & \multicolumn{3}{|c||}{\textbf{Inductive}} & \multicolumn{3}{|c|}{\textbf{Inductive + Transductive}} \\
& \textbf{Model} & SoundCloud & YouTube & Twitter & SoundCloud & YouTube & Twitter \\
\hline
\parbox[t]{3mm}{\multirow{8}{*}{\rotatebox[origin=c]{90}{Limited}}}
& Independent & 0.396 & 0.148 & 0.260 & 0.352 & 0.387 & 0.466 \\
& SGL(1) & 0.396 & 0.221 & 0.281 & 0.444 & 0.453 & 0.494 \\
& SGL(2) & 0.342 & 0.263 & 0.275 & 0.364 & 0.478 & 0.612 \\
& PSL & 0.363 & 0.225 & 0.249 & 0.332 & 0.438 & 0.498 \\
& MRF & \textbf{0.431} & 0.164 & 0.264 & 0.547 & 0.404 & 0.483 \\
& SGL(1) + PSL & 0.284 & 0.268 & 0.258 & 0.521 & 0.471 & 0.523 \\
& SGL(1) + MRF & 0.412 & 0.239 & \textbf{0.285} & 0.578 & 0.473 &  0.513 \\
& SGL(2) + PSL & 0.243 & \textbf{0.282} & 0.270 & 0.562 & 0.489 & 0.594 \\
& SGL(2) + MRF & 0.404 & 0.276 & 0.280 & \textbf{0.583} & \textbf{0.493} & \textbf{0.630} \\
\hline 
\hline
\parbox[t]{3mm}{\multirow{8}{*}{\rotatebox[origin=c]{90}{Full}}}
& Independent & 0.409 & 0.321 & \textbf{0.860} & 0.460 & 0.539 & \textbf{0.950} \\
& SGL(1) & 0.434 & \textbf{0.440} & 0.848 & 0.530 & 0.614 & 0.946 \\
& SGL(2) & 0.369 & 0.439 & 0.838 & 0.483 & \textbf{0.617} & 0.942 \\
& PSL & 0.458 & 0.337 & 0.823 & 0.469 & 0.569 & 0.927 \\
& MRF & 0.487 & 0.347 & 0.857 & 0.579 & 0.557 & 0.948 \\
& SGL(1) + PSL & 0.289 & 0.433 & 0.817 & 0.602 & 0.588 & 0.925 \\
& SGL(1) + MRF & \textbf{0.489} & 0.439 & 0.845 & \textbf{0.604} & 0.609 & 0.944 \\
& SGL(2) + PSL & 0.324 & 0.426 & 0.813 & 0.582 & 0.581 & 0.923 \\
& SGL(2) + MRF & 0.435 & 0.438 & 0.835 & \textbf{0.604} & 0.614 & 0.939 \\
\hline
\end{tabular}}
\end{center}
\end{table*}

\subsection*{Independent Model Performance}\label{sec:independent_performance}

We use logistic regression as our independent classifier, since it generally outperformed other methods such as tree ensembles during development.
Before testing our relational methods, we perform ablation tests on the independent model to determine which feature sets contribute the most. We test each feature set in isolation and in combination with each other on the first~10\% of the data for each domain using~70\% for training and tuning, and the remaining~30\% for testing~(Figure~\ref{fig:ablation}). The effectiveness of different feature types varies from domain to domain. For SoundCloud, a combination of content, graph, and user features (c+g+u) yields the best results; n-gram features are not very helpful. For YouTube and Twitter, the opposite is true: n-gram features on their own work almost as well as all feature types put together.

\subsection*{Relational Modeling Performance}\label{sec:relational_performance}

\begin{figure}[t]
  \mbox{\includegraphics[width=0.29\textwidth]{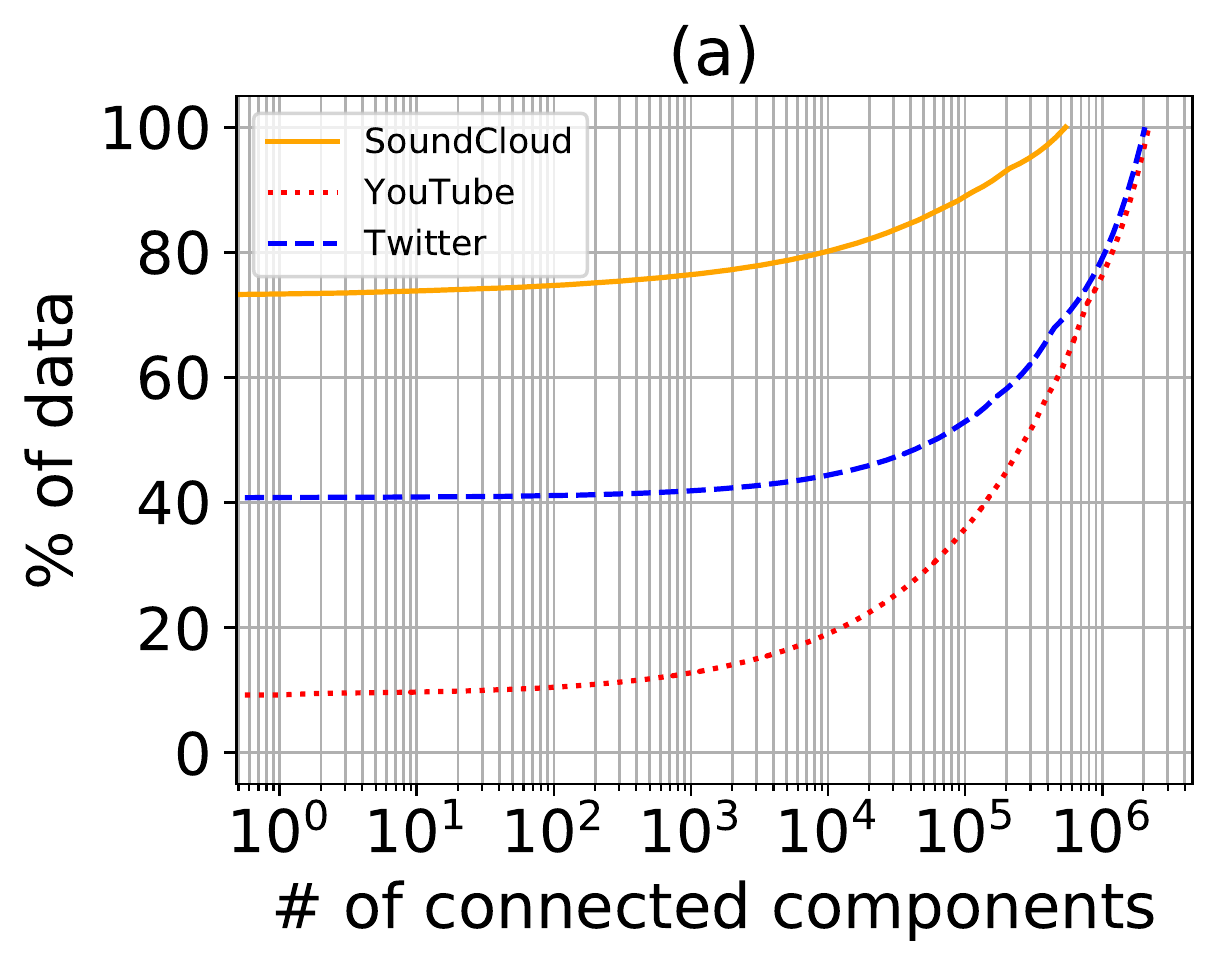}}
  \mbox{\includegraphics[width=0.17\textwidth]{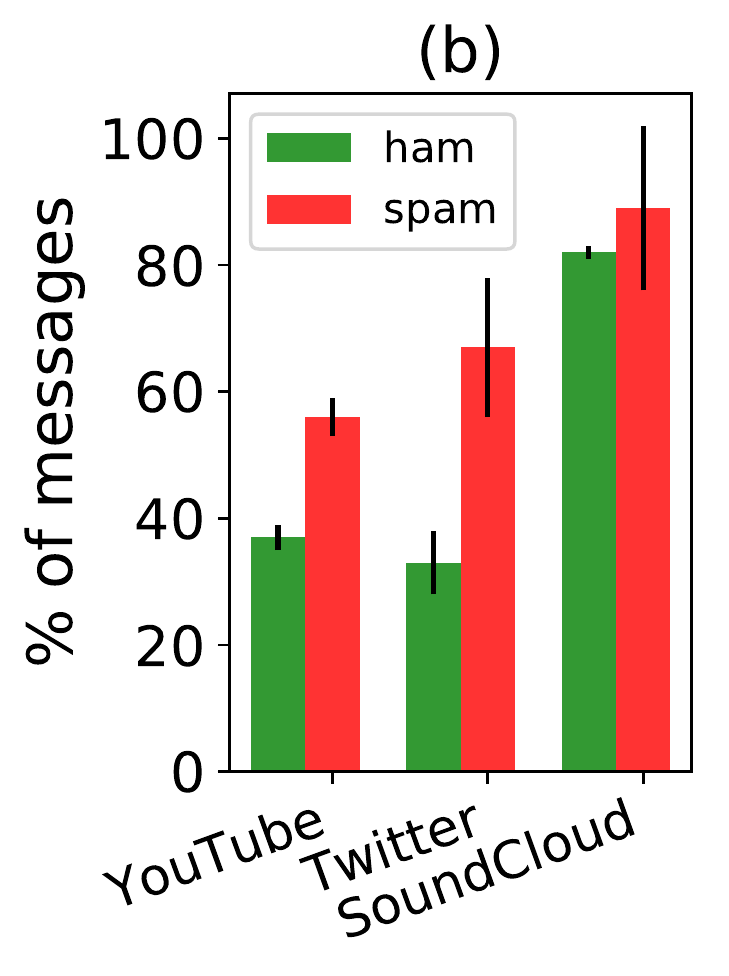}}
  \caption{(a) The number of connected components it takes to cover the first 5M messages for each domain. (b) Percentage of ham/spam test set messages with at least one connection to any training messages.}
  \label{fig:connected_components}
\end{figure}

We perform four sets of experiments to test our relational models' capability of detecting spam. The `full' experiments use all available features when learning and making predictions. The `limited' experiments exclude some of the most informative features: we remove n-gram features from YouTube and Twitter, and graph features from SoundCloud. This makes the prediction problem more challenging, testing the ability of our relational methods to compensate for limited information. In adversarial settings like spam, it's common to only have limited or noisy information, since spammers quickly adapt their messages and networks to evade detection.

We also evaluate our models for test instances that have no relational connections to any messages in the training sets (Inductive), in addition to evaluating all test instances (Inductive + Transductive). For SoundCloud, a small number of connected components covers a significant portion of ham and spam messages~(Figure~\ref{fig:connected_components}), but this is less so for YouTube and Twitter, where transductive methods would need more and more known labels to propagate to the increasingly disconnected pockets of related messages.

In all settings, we compare against predictions from the independent model (top row). Then we start to incorporate relational structure, applying our joint inference models to the outputs of the independent model as well as to the relationally augmented independent models, which use up to two levels of stacked learning.

 We find that not only can relational modeling improve performance for these domains~(Table~\ref{tab:overall_performance}), but in many cases stacked learning and joint reasoning work well together, achieving the best performance in many of the experiments. In the case of the limited feature set, relational modeling is able to improve performance for all three domains; in the fully featured case, relational modeling improves performance for the SoundCloud and YouTube domains. For Twitter, the independent model is already very effective when trained with n-grams (AUPR=0.95); stacked learning and joint inference offer no additional benefit. Against an evasive adversary that changes text to avoid detection, the independent model would be less effective and relational modeling more likely to help.

Comparing the different relational methods, we find that no one method dominates; thus, when applying EGGS to a new spam domain, we recommend testing several methods on validation data before committing to a single model.
Having more layers in stacking sometimes helps a lot, especially in the Inductive + Transductive setting. For example, on the limited Twitter dataset, SGL goes from 0.494 to 0.612 AUPR with the addition of a second layer, SGL+PSL goes from 0.523 to 0.594, and SGL+MRF goes from 0.513 to 0.630. In other cases, such as the full SoundCloud dataset in the Inductive setting, performance can drop: SGL goes from 0.434 to 0.369, and SGL+MRF goes from 0.489 to 0.435. With more layers, each classifier in SGL is trained on less data, which may explain why performance sometimes decreases.

MRF usually achieves better AUPR than PSL: out of the 12 combinations of datasets and inductive or transductive settings, MRF outperforms PSL on 8, SGL(1)+MRF outperforms SGL(1)+PSL on 10, and SGL(2)+MRF outperforms SGL(2)+PSL on 11.\footnote{This would be statistically significant under a binomial test (29 successes out of 36 trials), but the trials are not independent.} However, PSL usually achieves better AUROC, outperforming the MRF models in 25 out of 36 cases.

\section{Conclusion}\label{sec:conclusion}

We have shown how to flexibly incorporate relational structure from multiple separate domains using several approaches to build EGGS, a spam detection system that attacks the problem from as many angles as possible. Two of these approaches share a general relational framework that provides full joint inference over related messages, while the pseudo-relational features can be added to any independent model without the need for more complexity. We see that these methods are effective on real-world large-scale datasets in isolation and in combination with one another. Furthermore, if these specific techniques are a poor fit to a new domain, EGGS can easily be adapted to include other features, other types of classifiers, and other relational reasoning methods.

\section{Acknowledgements}

This research was supported by ARO grant W911NF-15-1-0265.

\bibliographystyle{aaai}
\bibliography{references}

\end{document}